\documentclass[12pt, a4paper]{article}
\usepackage{graphicx}
\usepackage{amssymb}
\usepackage{amsmath}
\usepackage{bm}
\usepackage{color}
\usepackage{theorem}
\usepackage{subcaption}
\usepackage{listings}
\usepackage{colortbl}
\usepackage{tabularx}
\usepackage{longtable}
\usepackage{url}
\usepackage[utf8]{inputenc}
\usepackage[T1]{fontenc}
\usepackage{lmodern}
\usepackage[top=30truemm,bottom=30truemm,left=25truemm,right=25truemm]{geometry}

\usepackage[sort&compress,numbers, merge]{natbib}

\definecolor{Orange}{cmyk}{0,0.61,0.87,0}
\definecolor{JungleGreen}{cmyk}{0.99,0,0.52,0}
\definecolor{OliveGreen}{cmyk}{0.64,0,0.95,0.40}
\definecolor{Brown}{cmyk}{0,0.81,1,0.60}
\definecolor{RoyalBlue}{cmyk}{0.71,0.53,0,0.12}
\definecolor{Gray}{cmyk}{0,0,0,0.40}
\definecolor{LightPink}{cmyk}{0.0,0.25,0,0}
\definecolor{LLightPink}{cmyk}{0.0,0.10,0,0}
\definecolor{LightBlue}{cmyk}{0.25,0,0,0}
\definecolor{LightGray}{cmyk}{0,0,0,0.2}

\allowdisplaybreaks[1]

\usepackage[colorlinks=true, linkcolor=OliveGreen, citecolor=RoyalBlue,
urlcolor=RoyalBlue]{hyperref}

\renewcommand{\thefootnote}{\fnsymbol{footnote}}

\begin{document}

\begin{titlepage}

  \begin{flushright}
    {\tt
      
    }
\end{flushright}

\vskip 1.35cm
\begin{center}

%\textcolor{RoyalBlue}
{\large
{\bf
Atomic Form Factors and Inverse Primakoff Scattering of Axion
}
}

\vskip 1.5cm

Tomohiro Abe$^{a}$\footnote{
\href{mailto:abetomo@hep-th.phys.s.u-tokyo.ac.jp}{\tt
abetomo@hep-th.phys.s.u-tokyo.ac.jp}},
Koichi Hamaguchi$^{a,b}$\footnote{
\href{mailto:hama@hep-th.phys.s.u-tokyo.ac.jp}{\tt
 hama@hep-th.phys.s.u-tokyo.ac.jp}},
and 
Natsumi Nagata$^a$\footnote{
\href{mailto:natsumi@hep-th.phys.s.u-tokyo.ac.jp}{\tt natsumi@hep-th.phys.s.u-tokyo.ac.jp}}

\vskip 0.8cm

{\it $^a$Department of Physics, University of Tokyo, Bunkyo-ku, Tokyo
 113--0033, Japan} \\[2pt]
{\it ${}^b$Kavli Institute for the Physics and Mathematics of the
 Universe (Kavli IPMU), University of Tokyo, Kashiwa 277--8583, Japan} 

\date{\today}

\vskip 1.5cm

\begin{abstract}
We reexamine the inverse Primakoff scattering of axions, whose scattering cross section depends on the distribution of electrons in target atoms. We evaluate it using a form factor computed with a relativistic Hartree-Fock wave function and compare it with the previous results obtained with those based on the screened Coulomb potential for the electrostatic field in the atom. We take xenon as an example for the target atom and show that the scattering cross section was overestimated by more than an order of magnitude for axions with 
%$\lesssim {\cal O}(1)$~keV 
$\lesssim {\cal O}(10)$~keV 
energies, like solar axions. It is also found that inelastic scattering processes, in which the final state contains an excited or ionized atom, can be comparable or even be dominant when the size of momentum transfer is $\lesssim 1$~keV. For more energetic axions, on the other hand, the total scattering cross section is found to be well approximated by a simple expression and has little dependence on the atomic structure. As an application of this result, we consider supernova axions, whose energy is 10--100~MeV, and show that $\mathcal{O}(1)$ inverse Primakoff events are expected for axions from a nearby supernova in the future neutrino experiments, which may warrant a more detailed study on the search strategy of this process.

\end{abstract}

\end{center}
\end{titlepage}

\renewcommand{\thefootnote}{\arabic{footnote}}
\setcounter{footnote}{0}

%%%%%%%%%%%%%%%%%%%%%%%%%%%%%%%%%%%%%%%%%%
\section{Introduction}
%%%%%%%%%%%%%%%%%%%%%%%%%%%%%%%%%%%%%%%%%%

A pseudo Nambu-Goldstone boson that has a coupling with the electromagnetic field is often predicted in physics beyond the Standard Model. The most famous candidate for such a particle is the axion~\cite{Weinberg:1977ma, Wilczek:1977pj}, which is associated with the spontaneous breaking of the Peccei-Quinn
symmetry~\cite{Peccei:1977hh, Peccei:1977ur} and couples to photons via the anomaly of this global 
symmetry. A great deal of experimental effort has been devoted to the search of such particles using their couplings with photons, having imposed stringent limits on the interaction strength. See Refs.~\cite{Irastorza:2018dyq, Armengaud:2019uso, DiLuzio:2020wdo, Sikivie:2020zpn, Zyla:2020zbs} for recent reviews on this class of particles, which we collectively call axion-like particles in what follows, as well as the latest status of their search results.

Many experiments probe axion-like particles through the so-called inverse Primakoff scattering process with an atom, in which the incoming axion-like particle is converted into a photon through the electromagnetic field generated by the atom. For example, axions produced in the Sun were searched for via this process with the help of the enhancement by the Bragg scattering~\cite{Paschos:1993yf, Avignone:1997th, Bernabei:2001ny, Morales:2001we, Ahmed:2009ht, Armengaud:2013rta}. Recently, this process has received renewed attention since it was pointed out in Refs.~\cite{Gao:2020wer, Dent:2020jhf} that the inverse Primakoff process might play an important role in the explanation of the excessive events observed in the XENON1T experiment~\cite{Aprile:2020tmw}. It is envisioned that this potential explanation will be tested in the near future~\cite{Aprile:2020vtw}.

In this paper, we revisit the calculation of the cross section of this inverse Primakoff scattering process, with particular attention to its dependence on atomic form factors. This dependence is most significant for axion-like particles with energies $\lesssim \mathcal{O}(1)$~keV, like solar axions, whose wavelength is similar to or larger than the atomic size. In this case, the resultant scattering cross section depends highly on the extent of screening of the internal electric field, which is determined by the distribution of electrons inside the target atom and taken into account in terms of an atomic form factor. In one of the earliest studies on this process, a simple Gaussian form of the atomic electron distribution was assumed, with a scale parameter fixed to be 1~{\AA} and 1.75~{\AA} for Ge and Pb, respectively~\cite{Avignone:1988bv}. In a later work~\cite{Buchmuller:1989rb}, on the other hand, the authors utilized a screened Coulomb potential for the electrostatic field in the target atom, which has the form of the Yukawa potential with the screening length $r_0$. This approach---and/or the resultant formula based on this approach---has been widely adopted in subsequent studies~\cite{Paschos:1993yf, Avignone:1997th, Creswick:1997pg, Cebrian:1998mu, Bernabei:2001ny, Morales:2001we, Ahmed:2009ht, Armengaud:2013rta,  Gao:2020wer, Dent:2020jhf}.\footnote{An exception was the calculation performed in Ref.~\cite{Li:2015tsa}, in which the charge density distribution for TeO$_2$ crystal was directly computed based on the density functional theory~\cite{Hohenberg:1964zz, Kohn:1965zzb}. } There are, however, several shortcomings in these methods. First and foremost, the distribution of electrons in an actual atom is more complicated than those assumed in these methods, and this difference may affect the resultant scattering cross section. In addition, the aforementioned approaches introduce a parameter with the dimension of length that needs to be fixed for each target atom, such as the scale parameter in Ref.~\cite{Avignone:1988bv} and the screening length $r_0$ in Ref.~\cite{Buchmuller:1989rb}; the determination of these parameters may cause uncertainty. For example, in Ref.~\cite{Dent:2020jhf}, the screening length for Xe was set to be equal to the Wigner-Seitz radius in liquid xenon, $r_0 = 2.45$~{\AA}, while in Ref.~\cite{Gao:2020wer}, it was determined through a fit of the assumed form of the form factor onto that obtained with the Hartree-Fock method, resulting in $r_0 = 0.49$~{\AA}.\footnote{A revised version of Ref.~\cite{Gao:2020wer}, Ref.~\cite{Gao:2020wer_v4}, used a new fitting with an extra parameter. See note added.}  This ambiguity in the parameter determination could potentially give rise to uncertainty in the calculation.  

To address these issues, in our calculation, we directly use a more realistic form factor that is computed with a relativistic Hartree-Fock wave function, without assuming a particular simple form for it. Taking xenon as an example for the target atom---motivated by the high sensitivities of the XENON experiments to solar axions---we show in Sec.~\ref{sec:elastic} that the cross section obtained with this form factor is actually smaller than the previous results by more than an order of magnitude. It is also found that the scattering cross section depends strongly on the choice of the screening length; an ${\cal O}(1)$ change in $r_0$ results in about an order of magnitude change in the scattering cross section. This strong dependence on the screening length, or on the ``atomic size'', is not specific to xenon but is a generic feature of the inverse Primakoff scattering with an atom, as we discuss below. It is, therefore, important to use a precise form factor for each target atom to calculate the cross section accurately. For convenience, in Appendix, we list approximated analytical functions for the form factors of several atoms that are used in experiments. 

Moreover, in Sec.~\ref{sec:inelastic}, we discuss inelastic scattering processes, in which the final state contains an excited or ionized atom; these processes are not considered in the previous studies. As it turns out, the cross sections of these processes are, in fact, comparable to the elastic one, or even be dominant especially for low-energy scatterings. The rate of the inelastic scattering contributing to the signal events depends on the actual experimental setup, and thus needs to be evaluated for each experiment. We encourage experimentalists to take account of the inelastic scattering processes as well in future experiments. 

The atomic form factor used in our analysis is computed for the size of momentum transfer $\lesssim 100$~keV. One may, thus, wonder how to compute the inverse Primakoff scattering rate for more energetic axions. To address this question, in Sec.~\ref{sec:snaxion}, we extend our consideration to the scattering of axions with energies much larger than ${\cal O}(1)$~keV. As we see, in this case, the total elastic scattering cross section is well approximated by a simple expression and has little dependence on the atomic size. Furthermore, the inelastic scattering cross section turns out to be always subdominant. We can therefore evaluate the inverse Primakoff rate without suffering from uncertainty coming from the atomic structure in this case. As an application of this result, we consider the scattering of supernova axions, whose typical energy is 10--100~MeV, and show that $\mathcal{O}(1)$ inverse Primakoff events are expected for axions from a nearby supernova in the future neutrino experiments, such as Hyper-Kamiokande~\cite{Abe:2018uyc}. This estimation encourages a more dedicated study on this search strategy for supernova axions.

%%%%%%%%%%%%%%%%%%%%%%%%%%%%%%%%%%%%%%%%%%%%%%
\section{Axion Inverse Primakoff Scattering}
\label{sec:elastic}
%%%%%%%%%%%%%%%%%%%%%%%%%%%%%%%%%%%%%%%%%%%%%%

%%%%%%%%%%%%%%%%%%%%%%%%%%%%%%%%%%%%%%%%%%%%%%%
\subsection{Formalism for elastic scattering}
\label{sec:formes}
%%%%%%%%%%%%%%%%%%%%%%%%%%%%%%%%%%%%%%%%%%%%%%%

We consider an axion-like particle $a$ which has a coupling to photons of the form 
\begin{equation}
  {\cal L}_{a\gamma\gamma} = \frac{g_{a\gamma\gamma}}{4} a F_{\mu\nu} \widetilde{F}^{\mu\nu} ~,
  \label{eq:lagg}
\end{equation}
where $\widetilde{F}_{\mu\nu} \equiv \frac{1}{2} \epsilon_{\mu\nu\rho\sigma} F^{\rho\sigma}$ with $\epsilon^{\mu\nu\rho\sigma}$ the totally antisymmetric tensor ($\epsilon^{0123} = +1$). We denote the mass of $a$ by $m_a$. For a QCD axion~\cite{Weinberg:1977ma, Wilczek:1977pj}, the axion mass $m_a$ is related to the axion decay constant $f_a$, which corresponds to the breaking scale of the Peccei-Quinn symmetry; at the next-to-next-to-leading order in chiral perturbation theory, we have~\cite{Gorghetto:2018ocs} 
\begin{equation}
  m_a = 5.691(51) \times \biggl(\frac{10^9~{\mathrm{GeV}}}{f_a}\biggr)~\mathrm{meV} ~. 
\end{equation}
The axion-photon coupling, $g_{a\gamma\gamma}$, is also related to the axion decay constant for a QCD axion. The relation depends on models and is given by~\cite{diCortona:2015ldu}
\begin{equation}
  g_{a\gamma\gamma} = \frac{\alpha}{2\pi f_a} \biggl[\frac{E}{N} - 1.92(4)\biggr] ~, 
\end{equation}
where $\alpha$ is the fine-structure constant and $E/N$ is the ratio between the electromagnetic and color anomaly factors for the Peccei-Quinn current; for instance, $E/N = 8/3$ for the DFSZ model~\cite{Zhitnitsky:1980tq, Dine:1981rt} while $E/N = 0$ for the KSVZ model~\cite{Kim:1979if, Shifman:1979if} with electrically neutral heavy fermions. In the following discussion, we do not restrict ourselves to these models, and regard both $m_a$ and $g_{a\gamma\gamma}$ as free parameters, as in the case of a generic axion-like particle.

The interaction~\eqref{eq:lagg} induces the axion-photon conversion through the scattering with an atom, $A$: $a + A \to \gamma + A$. This process is called the inverse Primakoff process and has been discussed in the literature~\cite{Avignone:1988bv, Buchmuller:1989rb, Paschos:1993yf, Avignone:1997th, Creswick:1997pg, Cebrian:1998mu, Bernabei:2001ny, Morales:2001we, Ahmed:2009ht, Armengaud:2013rta, Li:2015tsa, Gao:2020wer, Dent:2020jhf}, especially to search for solar axions. We also consider solar axions, whose typical energy is 
%${\cal O}(1)$~keV
a few keV,\footnote{For the energy spectrum of solar axions, see, \textit{e.g.}, Appendix~A in Ref.~\cite{Dent:2020jhf}. } as the prime target in the present and subsequent sections. In this section, we focus on the case in which the state of the atom is left unchanged during the scattering process, \textit{i.e.}, the scattering is elastic.

The differential cross section of this scattering process is 
\begin{equation}
  \frac{d\sigma_{\rm el}}{d\Omega} = 
  \frac{g_{a\gamma\gamma}^2 E_a^3 p_a \sin^2\theta}{16\pi^2 (E_a^2 + p_a^2 - 2 E_a p_a \cos \theta)^2 }
  \biggl| \int d^3 \bm{x} \, e^{-i \bm{q}\cdot \bm{x} } \langle A_0 | \rho (\bm{x} ) |A_0\rangle \biggr|^2 ~,
  \label{eq:dsigel0}
\end{equation}
where $E_a$ is the energy of the incoming axion, $\bm{q} \equiv \bm{p}_\gamma - \bm{p}_a$ with $\bm{p}_\gamma$ and $\bm{p}_a$ the momenta of the out-going photon and incoming axion, respectively, $p_a \equiv |\bm{p}_a|$, $\theta$ is the scattering angle, $\rho (\bm{x})$ is the charge density operator, and $|A_0\rangle$ denotes the ground state of the atom $A$ with $\langle A_0|A_0\rangle = 1$. We have taken the non-relativistic limit for the atomic state $|A_0\rangle$, with which the matrix element of the current operator $\bm{J}(\bm{x})$ vanishes and the photon energy is equal to the axion energy $E_a$. We also note that $\bm{q}^2 = E_a^2 + p_a^2 - 2 E_a p_a \cos \theta$. In an atom of atomic number $Z$, the charge density $\rho (\bm{x})$ can be written as 
\begin{align}
  \rho (\bm{x}) & = Z e\delta^3 (\bm{x}) - e n_e (\bm{x}) \notag \\
  & = Z e\delta^3 (\bm{x}) - e \sum_i \delta^3 (\bm{x}-\bm{x}_i) ~,
  \label{eq:rho}
\end{align}
where $e > 0$ denotes the positron charge, $n_e (\bm{x})$ is the electron number density, and $\bm{x}_i$ is the position of each atomic electron, with the origin taken at the nucleus of the atom. By substituting this into Eq.~\eqref{eq:dsigel0}, we obtain 
\begin{equation}
  \frac{d\sigma_{\rm el}}{d\Omega} = 
  \frac{\alpha g_{a\gamma\gamma}^2 E_a^3 p_a \sin^2\theta}{4\pi (E_a^2 + p_a^2 - 2 E_a p_a \cos \theta)^2 }
  \bigl| Z - F(\bm{q}) \bigr|^2 ~,
  \label{eq:dsigel1}
\end{equation}
where $F(\bm{q})$ is the atomic form factor: 
\begin{align}
  F(\bm{q}) &\equiv \int d^3 \bm{x} \, e^{-i \bm{q}\cdot \bm{x} } n_e (\bm{x}) = \sum_i  \langle A_0 | e^{-i \bm{q}\cdot \bm{x}_i } |A_0\rangle ~. 
  \label{eq:formfactor}
\end{align}
In particular, for $m_a \ll E_a$, we have 
\begin{equation}
  \frac{d\sigma_{\rm el}}{d\Omega} \simeq
  \frac{\alpha g_{a\gamma\gamma}^2  \sin^2\theta}{16\pi (1-\cos\theta)^2}\bigl| Z - F(\bm{q}) \bigr|^2 ~.
  \label{eq:dsigelml}
\end{equation}

Notice that if the size of the momentum transfer $|\bm{q}|$ is much larger than the inverse of the atomic size, the integrand in Eq.~\eqref{eq:formfactor} is a rapidly oscillating function and thus the integral is highly suppressed; therefore, in this limit, $Z - F(\bm{q} ) \to Z$. On the other hand, for $|\bm{q}| \to 0$, we can expand $e^{-i \bm{q}\cdot \bm{x} }$ in terms of $\bm{q}\cdot \bm{x}$ in Eq.~\eqref{eq:formfactor} as
\begin{equation}
  F(\bm{q}) \simeq \int d^3 \bm{x} \, n_e (\bm{x}) \biggl[1 -\frac{1}{2} (\bm{q}\cdot \bm{x})^2 \biggr]
  = Z - \frac{1}{2}\int d^3 \bm{x} \, (\bm{q}\cdot \bm{x})^2  n_e(\bm{x}) 
 ~.
\end{equation}
If the distribution of electrons is spherically symmetric, the right-hand side leads to 
\begin{equation}
  Z -F(\bm{q}) \simeq \frac{\bm{q}^2}{6} Z \langle r^2 \rangle ~,
  \label{eq:fqexpand}
\end{equation}
where $\langle r^2 \rangle$ is the mean square radius of the atom: 
\begin{equation}
  \langle r^2 \rangle \equiv 
  \frac{\int d^3 \bm{x} \,  \bm{x}^2  n_e(\bm{x})}{\int d^3 \bm{x} \, n_e(\bm{x})} ~
  =
    \frac{\int d^3 \bm{x} \,  \bm{x}^2  n_e(\bm{x})}{Z} ~.
\end{equation}
In this case, the differential scattering cross section~\eqref{eq:dsigel1} has a simple form 
\begin{equation}
  \frac{d\sigma_{\rm el}}{d\Omega} \simeq 
  \frac{\alpha g_{a\gamma\gamma}^2 E_a^3 p_a }{144\pi } Z^2 \langle r^2 \rangle^2 \sin^2\theta ~,
  \label{eq:dsigellowenergy}
\end{equation}
and, in particular, vanishes in the forward region.\footnote{Notice that this is contrary to the case of the axion-photon conversion in a homogeneous static magnetic field, whose differential cross section peaks in the forward direction.  }

%%%%%%%%%%%%%%%%%%%%%%%%%%%%%%%%
\subsection{Atomic form factor}
\label{sec:formfactor}
%%%%%%%%%%%%%%%%%%%%%%%%%%%%%%%%

The atomic form factor $F(\bm{q})$ in Eq.~\eqref{eq:formfactor} needs to be evaluated for each target atom. To be specific, in this work, we consider a xenon $(Z=54)$ as the target atom.

An approach to estimate the form factor adopted in the literature is to assume a screened Coulomb potential for the electrostatic field in the atom~\cite{Buchmuller:1989rb, Cebrian:1998mu, Ahmed:2009ht}: 
\begin{equation}
  \phi (r) = \frac{Ze}{4\pi r} e^{-\frac{r}{r_0}} ~,
  \label{eq:yukawapot}
\end{equation}
where $r$ is the distance from the nucleus and $r_0$ is the screening length. By noting that the potential $\phi$ is related to the charge density via Poisson's equation, 
\begin{equation}
  \nabla^2 \phi = - \rho ~,
\end{equation}
we see that the integral in Eq.~\eqref{eq:dsigel0} can be evaluated as\footnote{$F_a (2 \Theta)$ in Ref.~\cite{Buchmuller:1989rb} is related to this integral by 
\begin{equation}
  F_a (2 \Theta) = \frac{k^2}{q^2}  \int d^3 \bm{x} \, e^{-i \bm{q}\cdot \bm{x} } \rho (\bm{x} ) ~,
\end{equation}
with $k = |\bm{p}_{\gamma}|$ and $\Theta = \theta/2$. 
} 
\begin{equation}
  \int d^3 \bm{x} \, e^{-i \bm{q}\cdot \bm{x} } \rho (\bm{x} ) 
  = \frac{Zeq^2}{q^2 + (1/r_0)^2} ~,
\end{equation}
with $q \equiv |\bm{q}|$. As a result, from Eqs.~\eqref{eq:rho} and \eqref{eq:formfactor} we obtain a form factor of the form 
\begin{equation}
  F_{\mathrm{sc}} (q; r_0) = \frac{Z}{1 + q^2 r_0^2} ~.
  \label{eq:fsc}
\end{equation}
Using this form factor, we can readily evaluate the differential cross section from Eq.~\eqref{eq:dsigel1},
\begin{equation}
  \frac{d\sigma_{\rm el}}{d\Omega} \biggr|_{\mathrm{sc}} = 
  \frac{\alpha Z^2 g_{a\gamma\gamma}^2 r_0^4 E_a^3 p_a \sin^2\theta}{4\pi [1+r_0^2 (E_a^2 + p_a^2 - 2 E_a p_a \cos \theta) ]^2 } ~,
\end{equation}
and the total cross section 
\begin{equation}
  \sigma_{\mathrm{el}} \bigr|_{\mathrm{sc}}= 
  \frac{\alpha Z^2 g_{a\gamma\gamma}^2  }{2 } \frac{E_a}{p_a}
  \biggl[\frac{r_0^2(E_a^2 +p_a^2) +1 }{4r_0^2 E_a p_a  } \ln \biggl\{\frac{1 +r_0^2 (E_a + p_a)^2 }{1+r_0^2 (E_a-p_a)^2 }\biggr\} -1 \biggr] ~.
\end{equation}
For $m_a = 0$, these equations lead to 
\begin{equation}
  \frac{d\sigma_{\rm el}}{d\Omega}\biggr|_{\mathrm{sc}}  = \frac{\alpha Z^2 g_{a\gamma\gamma}^2 r_0^4 E_a^4 \sin^2\theta}{4\pi [1+ 2 E_a^2 r_0^2 (1-\cos \theta) ]^2} ~,
\end{equation}
and 
\begin{equation}
  \sigma_{\mathrm{el}}\bigr|_{\mathrm{sc}} = 
  \frac{\alpha Z^2 g_{a\gamma\gamma}^2  }{2 } 
  \biggl[\frac{2r_0^2E_a^2  +1 }{4r_0^2 E_a^2 } \ln (1 + 4 r_0^2 E_a^2) -1 \biggr]
  ~,
  \label{eq:previousformula}
\end{equation}
which agree with the results given in Refs.~\cite{Buchmuller:1989rb, Avignone:1997th, Creswick:1997pg, Gao:2020wer, Dent:2020jhf}.

%%%%%%%%%%%%%%%%%%%%%%%%%%%%%%%%%%%%%%%%%%%%%%%%%%%%%%%%%%%%%%%%%%%%
\begin{figure}
  \centering
  {\includegraphics[width=0.55\textwidth]{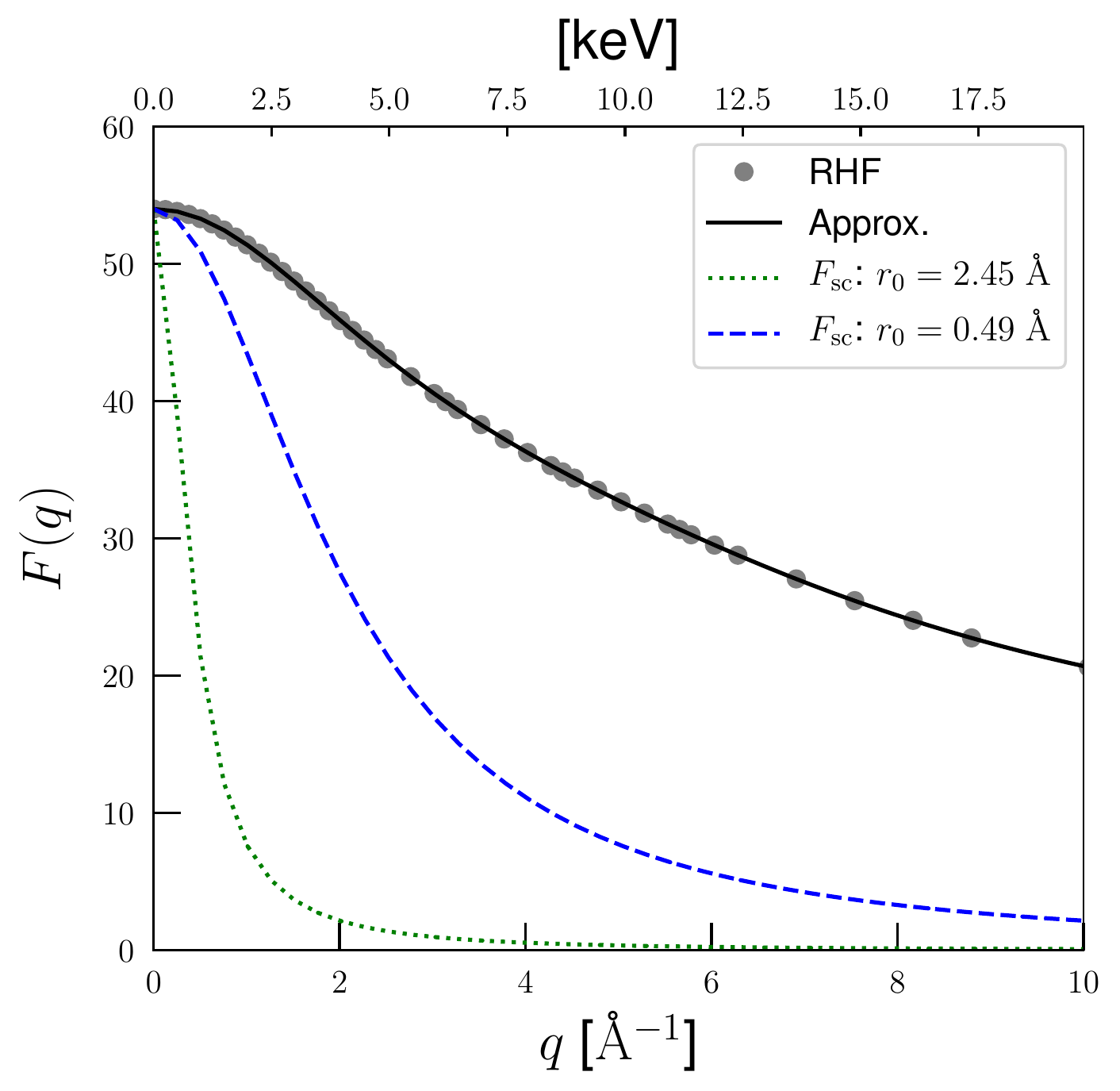}} 
  \caption{The atomic form factor of Xe as a function of $q \equiv |\bm{q}|$ obtained with the relativistic Hartree-Fock wave function (gray dots)~\cite{formfactor}, compared to those obtained from the screened Coulomb potential with $r_0 = 2.45$~{\AA} and 0.49~{\AA} in the green dotted and blue dashed lines, which are used in Ref.~\cite{Dent:2020jhf} and Ref.~\cite{Gao:2020wer}, respectively. The approximated function in Eq.~\eqref{eq:apprx} is also shown in the black solid line. 
  }
  \label{fig:ff}
  \end{figure}
  %%%%%%%%%%%%%%%%%%%%%%%%%%%%%%%%%%%%%%%%%%%%%%%%%%%%%%%%%%%%%%%%%%%%%

In a realistic atom, however, the electrostatic potential cannot be expressed in a simple form as in Eq.~\eqref{eq:yukawapot}, and thus the form factor does not have a form~\eqref{eq:fsc} either. We, therefore, need to compute it using a ground-state wave function for each atom, \textit{i.e.}, the state $|A_0\rangle$ in Eq.~\eqref{eq:formfactor}. A list of the results for such calculations can be found in Ref.~\cite{formfactor}, where the atomic form factor for Xe is obtained based on the calculation in Ref.~\cite{Doyle:a05916}, which utilizes a relativistic Hartree-Fock (RHF) wave function provided in Ref.~\cite{Coulthard_1967}. The resultant wave function is a function of $q \equiv |\bm{q}|$ and tabulated in Ref.~\cite{formfactor}.\footnote{Notice that it is, actually, tabulated with respect to $s \equiv |\bm{q}|/(4\pi)$ in units of {\AA}$^{-1}$. } 
In Fig.~\ref{fig:ff}, we plot the xenon atomic form factor provided in Ref.~\cite{formfactor} with gray blobs. We also show $F_{\mathrm{sc}}(q; r_0)$ with $r_0 = 2.45$~{\AA} and 0.49~{\AA} in the green dotted and blue dashed lines, which are used in Ref.~\cite{Dent:2020jhf} and Ref.~\cite{Gao:2020wer}, respectively. As we see, the form factor computed with the RHF wave function is larger than those used in Refs.~\cite{Gao:2020wer, Dent:2020jhf} for $|\bm{q}| \gtrsim 1$~\AA$^{-1}$; namely, electrons in the atom distribute rather close to the nucleus, compared with the distribution corresponding to a screened Coulomb potential. 

It is known that the atomic form factors obtained in Ref.~\cite{formfactor} are well approximated by the following function:\footnote{A similar analytical expression can be found in Ref.~\cite{Waasmaier}, which is implemented in the XrayDB Python module: \url{https://xraypy.github.io/XrayDB/}. We find that the difference between these two approximations is negligible in the range of $q$ shown in Fig.~\ref{fig:ff}. }
\begin{equation}
  F(q) \simeq \sum_{i = 1}^{4} a_i \exp \biggl[- b_i \biggl(\frac{|\bm{q}|}{4\pi}\biggr)^2\biggr] + c ~,
  \label{eq:apprx}
\end{equation}
where $|\bm{q}|$ is in the units of \AA$^{-1}$.
For Xe, the parameters $a_i$, $b_i$, and $c$ are obtained as follows~\cite{formfactor}: $a_1 = 20.2933$, $b_1 = 3.92820$, $a_2 = 19.0298$, $b_2 = 0.344000$, $a_3 = 8.97670$, $b_3 = 26.4659$, $a_4 = 1.99000$, $b_4 = 64.2658$, and $c = 3.71180$. We show the function in Eq.~\eqref{eq:apprx} with these parameters in the black solid line in Fig.~\ref{fig:ff}. It is found that the approximated function~\eqref{eq:apprx} offers an extremely good fit to the atomic form factor computed in Ref.~\cite{formfactor}. We, therefore, use this analytical expression to compute the scattering cross section in the next subsection.\footnote{We summarize the parameters of this analytical expression for a selected choice of other target atoms and ions in Table~\ref{tab:formfactors} in Appendix. }
The fitting function in Eq.~\eqref{eq:apprx} gives a close fit to the atomic form factor for $|\bm{q}|\lesssim 25$ \AA$^{-1}$~\cite{formfactor}. For a much larger $|\bm {q}|$, however, there should be no screening, i.e., $F(\bm {q})\to 0$, as discussed in Sec.~\ref{sec:formes}, and therefore Eq.~\eqref{eq:apprx} becomes in accurate. We discuss such a case in Sec.~\ref{sec:snaxion}.

%%%%%%%%%%%%%%%%%%%%%%%%%%%%
\subsection{Cross section}
%%%%%%%%%%%%%%%%%%%%%%%%%%%%

We have seen in the previous subsection that the form factor computed with a relativistic Hartree-Fock wave function considerably differs from those obtained with a screened Coulomb potential for $|\bm{q}| \gtrsim 1$~\AA$^{-1}$. This difference may have a significant impact on the scattering cross section of axions with energy of ${\cal O}(1)$~keV. We, in particular, note that the formula used in the previous works~\cite{Buchmuller:1989rb, Avignone:1997th, Creswick:1997pg, Cebrian:1998mu, Gao:2020wer, Dent:2020jhf} is based on the form factor of the form~\eqref{eq:fsc}, \textit{i.e.}, associated with a screened Coulomb potential, as we have seen above. In this subsection, we compute the cross section with the form factor obtained given in Ref.~\cite{formfactor} and compare it with those used in the previous works. 

To that end, we first obtain a formula for the total scattering cross section in a more useful form by noting that the form factor given in Ref.~\cite{formfactor} is a function of $q = |\bm{q} | $ and Eq.~\eqref{eq:dsigel1} can be integrated as 
\begin{equation}
  \sigma_{\mathrm{el}} = \frac{\alpha g_{a\gamma\gamma}^2}{8} \int_{E_a- p_a}^{E_a + p_a} dq \, \frac{[q^2 -(E_a - p_a)^2][(E_a + p_a)^2 -q^2]}{p_a^2 q^3}
  \bigl| Z - F(q) \bigr|^2 ~.
\end{equation}
We then compute this integral numerically.

%%%%%%%%%%%%%%%%%%%%%%%%%%%%%%%%%%%%%%%%%%%%%%%%%%%%%%%%%%%%%%%%%%%%
\begin{figure}
  \centering
  {\includegraphics[width=0.55\textwidth]{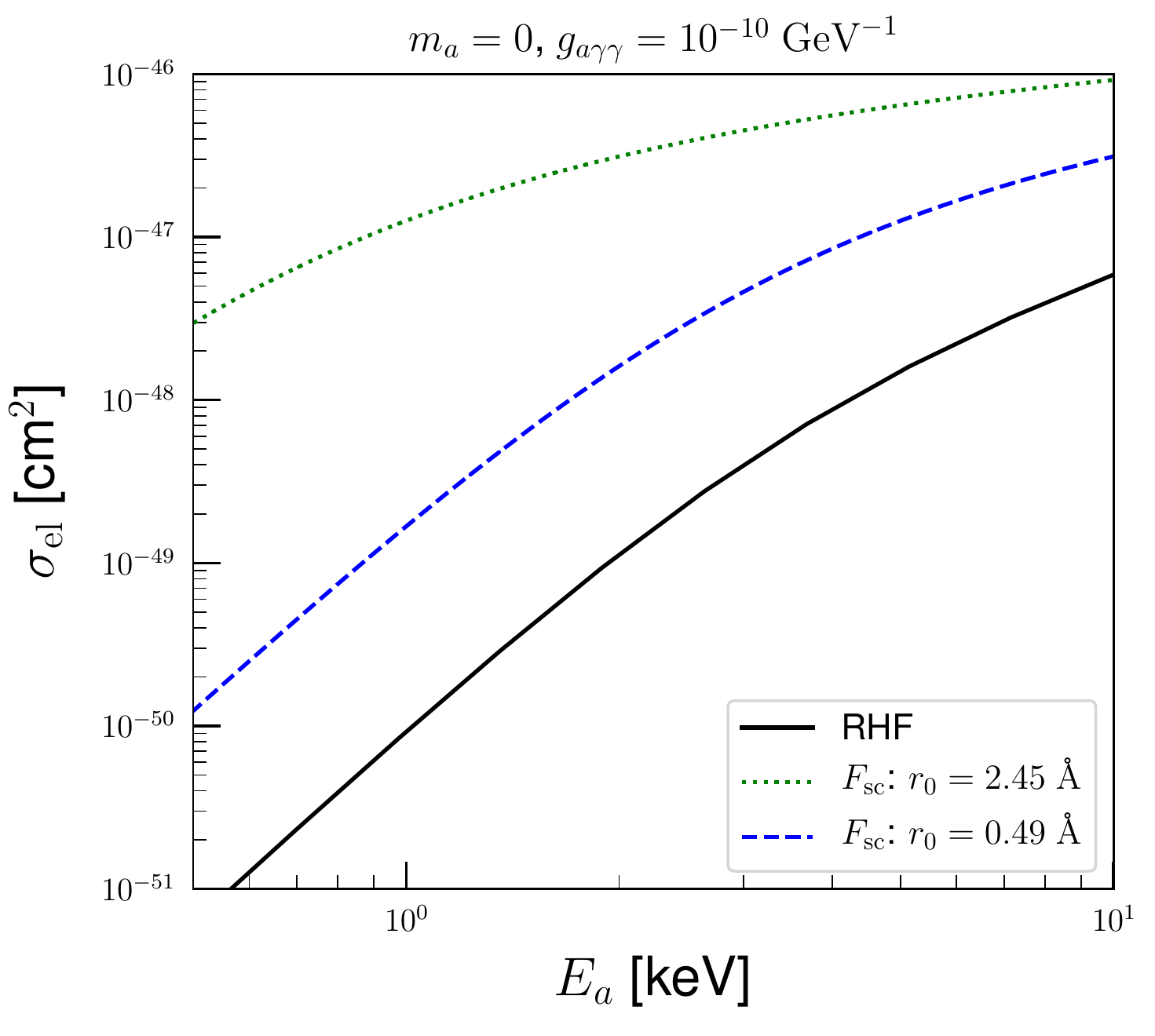}} 
  \caption{The total elastic scattering cross section of the inverse Primakoff process on Xe as a function of the axion energy $E_a$ with a realistic atomic form factor in the black solid line, compared to those obtained from a screened Coulomb potential with $r_0 = 2.45$~{\AA}~\cite{Dent:2020jhf} and 0.49~{\AA}~\cite{Gao:2020wer} in the green dotted and blue dashed lines, respectively. We set $m_a = 0$ and $g_{a\gamma\gamma} = 10^{-10}~\mathrm{GeV}^{-1}$.
  }
  \label{fig:elcross}
  \end{figure}
  %%%%%%%%%%%%%%%%%%%%%%%%%%%%%%%%%%%%%%%%%%%%%%%%%%%%%%%%%%%%%%%%%%%%%

In Fig.~\ref{fig:elcross}, we show the resultant total elastic scattering cross section $\sigma_{\rm el}$ on Xe as a function of the axion energy $E_a$ in the black solid line for $m_a = 0$ and $g_{a\gamma\gamma} = 10^{-10}~\mathrm{GeV}^{-1}$. As a comparison, we also show the results obtained with the form factor $F_{\rm sc}$ in Eq.~\eqref{eq:fsc} with $r_0 = 2.45$~{\AA} and 0.49~{\AA} in the green dotted and blue dashed lines, respectively, which correspond to the cross sections used in Ref.~\cite{Dent:2020jhf} and Ref.~\cite{Gao:2020wer}, respectively. It turns out that the cross section obtained in our analysis is smaller than those used in the previous studies by about an order of magnitude, or more---this smallness can be attributed to the larger value of $F(q)$ as we have seen in Fig.~\ref{fig:ff}. 

The size of the discrepancy among the results shown in Fig.~\ref{fig:elcross}, especially for $E_a \lesssim 1$~keV, can be understood in a quantitative manner as follows. Recall that for low-energy scatterings the cross section is proportional to the square of the mean square radius of the atom, $\langle r^2 \rangle$; from Eq.~\eqref{eq:dsigellowenergy}, we have
\begin{equation}
  \sigma_{\mathrm{el}} \simeq \frac{1}{54} \alpha g_{a\gamma\gamma}^2 E_a^3 p_a Z^2 \langle r^2 \rangle^2 ~,
\end{equation}
for low-energy scatterings.
The mean square radius of an atom with a screened Coulomb potential can readily be obtained from Eq.~\eqref{eq:fqexpand} and Eq.~\eqref{eq:fsc} as 
\begin{equation}
  \sqrt{\langle r^2 \rangle_{\mathrm{sc}} } = \sqrt{6} \, r_0
  \simeq 
  \begin{cases}
    6.0~\text{\AA} & (r_0 = 2.45~\text{\AA}) \\
    1.2 ~\text{\AA} & (r_0 = 0.49~\text{\AA})
  \end{cases} 
  ~.
  \label{eq:r2sc}
\end{equation}
On the other hand, from the approximated analytical formula~\eqref{eq:apprx}, we can evaluate the mean square radius for this atomic form factor as 
\begin{equation}
  \sqrt{\langle r^2 \rangle } = \frac{1}{4\pi} \sqrt{\frac{6}{Z} \sum_{i=1}^{4} a_i b_i} 
  \simeq 0.56~\text{\AA} ~.
  \label{eq:r2r}
\end{equation}
We see that if we use a screened Coulomb potential with $r_0 = 2.45$~{\AA} and 0.49~{\AA}, we overestimate $\sqrt{\langle r^2 \rangle }$ by a factor of $\simeq 11$ and $2.1$ and the scattering cross section by a factor of $1.2 \times 10^4$ and $21$, respectively---this accounts for the discrepancy at low energies in Fig.~\ref{fig:elcross}. 

All in all, our result indicates that it is important to take account of the correct structure of atoms---or, to use a precise value of the ``atomic size''---to accurately evaluate the cross section of the inverse Primakoff process for 
axions with $\lesssim {\cal O}(10)$~keV energies.
%keV axions.

%%%%%%%%%%%%%%%%%%%%%%%%%%%%%%%%
\section{Inelastic Scattering}
\label{sec:inelastic}
%%%%%%%%%%%%%%%%%%%%%%%%%%%%%%%%

Axions may excite the target atom in the inverse Primakoff process, which may be indistinguishable from the elastic scattering process in experiments. We discuss the significance of these processes compared to the elastic one in this section. 

Suppose that the target atom that was in the ground state $|A_0\rangle$ is excited into a state $|A_n\rangle$ in the scattering process; we denote the energies of these states by $E_0$ and $E_n$, respectively. The excited state may either be in the discrete or continuous spectrum. In the latter case, the atom is ionized after the scattering.

In order to obtain the inelastic scattering cross section, we need to know the energy spectrum $E_n$ as well as the matrix elements associated with the states $|A_n\rangle$ and $|A_0\rangle$, which requires the computation of the atomic wave functions for each state. Such a study is far beyond the scope of our analysis. Instead, we focus on the case where both the axion energy and the size of the transferred momentum is much larger than the difference in energy levels, $E_n - E_0$. This condition usually holds for $E_a = {\cal O}(1)$~keV. The differential scattering cross section under this condition can then be obtained in a similar manner as before. We have 
\begin{equation}
  \frac{d\sigma_{\rm inel}^{(n)}}{d\Omega} \simeq 
  \frac{g_{a\gamma\gamma}^2 E_a^3 p_a \sin^2\theta}{16\pi^2 (E_a^2 + p_a^2 - 2 E_a p_a \cos \theta)^2 }
  \biggl| \int d^3 \bm{x} \, e^{-i \bm{q}\cdot \bm{x} } \langle A_n | \rho (\bm{x} ) |A_0\rangle \biggr|^2 ~.
  \label{eq:dsiginel0}
\end{equation}
We evaluate the above integral using Eq.~\eqref{eq:rho} and noting that $\langle A_n|A_0 \rangle = 0$: 
\begin{align}
 \int d^3 \bm{x} \, e^{-i \bm{q}\cdot \bm{x} } \langle A_n | \rho (\bm{x} ) |A_0\rangle &= -e \sum_i
 \langle A_n | e^{-i \bm{q}\cdot \bm{x}_i } |A_0\rangle  ~.
\end{align}
We further define the following quantity, called the incoherent scattering function: 
\begin{equation}
  S (\bm{q}, Z) \equiv \sum_{n>0} \biggl| \sum_i \langle A_n |  e^{-i \bm{q}\cdot \bm{x}_i } |A_0\rangle  \biggr|^2~,
  \label{eq:incscf}
\end{equation}
where the sum is taken over all states (both the discrete and continuum states) except for the ground state ($n = 0$). With this function, we can express the sum of Eq.~\eqref{eq:dsiginel0} over $n > 0$ as 
\begin{equation}
  \sum_n \frac{d\sigma_{\rm inel}^{(n)}}{d\Omega} \simeq 
  \frac{\alpha g_{a\gamma\gamma}^2 E_a^3 p_a \sin^2\theta}{4\pi (E_a^2 + p_a^2 - 2 E_a p_a \cos \theta)^2 } S (\bm{q}, Z) ~. 
\end{equation}
By comparing this expression with Eq.~\eqref{eq:dsigel1}, we see that the significance of inelastic scattering processes can be estimated by comparing the incoherent scattering function $S(\bm{q},Z)$ with the factor $|Z -F(\bm{q})|^2$.

%%%%%%%%%%%%%%%%%%%%%%%%%%%%%%%%%%%%%%%%%%%%%%%%%%%%%%%%%%%%%%%%%%%%
\begin{figure}
  \centering
  {\includegraphics[width=0.55\textwidth]{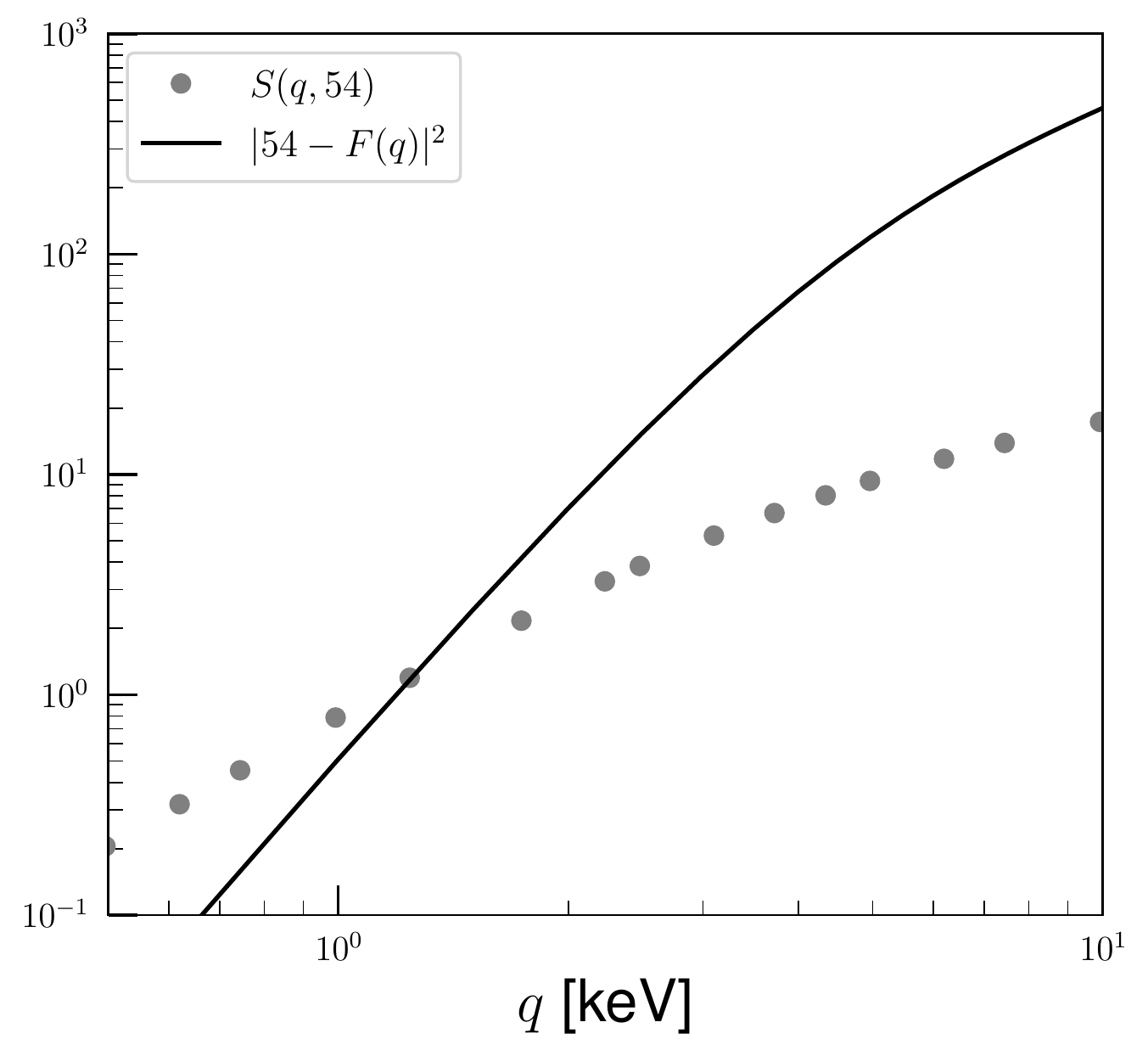}} 
  \caption{ The incoherent scattering function $S(q, Z)$ and the factor $|Z -F(q)|^2$ against the size of the transferred momentum $q=|\bm{q}| $ for xenon ($Z = 54$) in the gray dots and black solid line, respectively. We take the values of $S(q, Z)$ from Ref.~\cite{doi:10.1063/1.555523} and use the approximated expression in Eq.~\eqref{eq:apprx} for $|Z -F(q)|^2$. 
  }
  \label{fig:inchsf}
  \end{figure}
  %%%%%%%%%%%%%%%%%%%%%%%%%%%%%%%%%%%%%%%%%%%%%%%%%%%%%%%%%%%%%%%%%%%%%

In Fig.~\ref{fig:inchsf}, we plot the incoherent scattering function $S(q, Z)$ and the factor $|Z -F(q)|^2$ against the size of the transferred momentum $q=|\bm{q}| $ for xenon ($Z = 54$) in the gray dots and black solid line, respectively. We take the values of $S(q, Z)$ from Ref.~\cite{doi:10.1063/1.555523} and use the approximated expression in Eq.~\eqref{eq:apprx} for $|Z -F(q)|^2$. From this plot, we see that the inelastic scattering processes can be important for the keV-scale momentum exchange and even become dominant for $q \lesssim 1$~keV. 

In a realistic experimental setup, however, some of the inelastic scattering processes may not contribute to the signal. For instance, when the energy transfer is rather large, the photon energy in the final state, which is equal to $E_a -(E_n - E_0)$, may not exceed the energy threshold. We also note that when the energy transfer is sizable, it is inappropriate to use our approximated formula~\eqref{eq:dsiginel0}. To obtain a more precise expression for the differential cross section in this case, we need to compute the energy levels $E_n$ and the corresponding matrix elements of the electromagnetic current; to that end, we may use wave functions obtained with, \textit{e.g.}, the flexible atomic code~\cite{doi:10.1139/p07-197}, as done in Ref.~\cite{Ibe:2017yqa}. The total inelastic scattering cross section is then obtained by summing the cross sections for all of the channels that satisfy the signal selection criteria, with efficiency factors taken into account. Such a computation needs to be done for each experimental setup and, of course, beyond the scope of this paper---we leave this for future work.

%%%%%%%%%%%%%%%%%%%%%%%%%%%%%%%%%%%%%%%%%%%%%%%%%%%%%%%%%%%%%
\section{Inverse Primakoff Scattering for Supernova Axions}
\label{sec:snaxion}
%%%%%%%%%%%%%%%%%%%%%%%%%%%%%%%%%%%%%%%%%%%%%%%%%%%%%%%%%%%%%

Another interesting target for the energy range of axions is 10--100~MeV, a typical energy of supernova axions. A huge amount of axions can be produced from a proto-neutron star generated in core-collapse supernova explosion. Indeed, one of the most stringent limits on axion models\footnote{The temperature observations of neutron stars give independent limits on axion models~\cite{Hamaguchi:2018oqw, Beznogov:2018fda, Leinson:2019cqv}, which are as strong as the SN1987A limit. } is set by SN1987A~\cite{Ellis:1987pk, Raffelt:1987yt, Turner:1987by, Mayle:1987as, 
Brinkmann:1988vi, Burrows:1988ah, Mayle:1989yx, Chang:2018rso,
Carenza:2019pxu, Carenza:2020cis}, where the bound is obtained from the observed duration of the neutrino events~\cite{Hirata:1987hu, Bionta:1987qt, Alekseev:1987ej} as the axion emission tends to shorten the duration of neutrino emission. If a supernova explosion occurs near the earth in the future, we may be able to detect axions from the supernova~\cite{Raffelt:2011ft, Meyer:2016wrm, Irastorza:2018dyq, Carenza:2018jjc, Carenza:2018vcb, Meyer:2020vzy, Ge:2020zww}; this motivates us to consider the inverse Primakoff scattering of axions with 10--100~MeV energies. 

Let $r_A$ be the typical size of the target atom. For $E_a > \mathcal{O}(10)$~MeV, $|\bm{q}| r_A \gg 1$ unless the scattering angle is extremely small. For $|\bm{q}| r_A \gg 1$, as discussed in Sec.~\ref{sec:formes}, we can safely set $Z - F(\bm{q})$ to be $Z$, \textit{i.e.}, $F(\bm{q}) = 0$. On the other hand, as seen in Eq.~\eqref{eq:dsigellowenergy}, the differential scattering cross section is suppressed for small scattering angles, and thus we can neglect the contribution of the forward scattering to the total cross section. As a result, the total scattering cross section in this case can be expressed in a simple form
\begin{equation}
  \sigma_{\mathrm{el}} \simeq \frac{\alpha Z^2 g_{a\gamma\gamma}^2}{8 E_a^2} \int_{r_A^{-1}}^{2E_a } dq \, \frac{[4 E_a^2 -q^2]}{ q}
  \simeq \frac{\alpha Z^2 g_{a\gamma\gamma}^2}{2} 
  \biggl[\ln (2 E_a r_A)- \frac{1}{2}\biggr] 
  ~,
  \label{eq:sigelhe}
\end{equation}
where we set $m_a = 0$ just for simplicity and take the lower limit of integration to be the lower bound of the condition $qr_A > 1$. We obtain a similar expression from Eq.~\eqref{eq:previousformula} in the limit $E_a r_0 \gg 1$: 
\begin{equation}
  \sigma_{\mathrm{el}} \simeq \frac{\alpha Z^2 g_{a\gamma\gamma}^2}{2} 
  \bigl[\ln (2 E_a r_0)- 1\bigr]  ~.
\end{equation}
As we see from these equations, the total scattering cross section depends on the atomic size only logarithmically, and thus for a large $E_a$ we do not need detailed knowledge on the atomic structure. For instance, for $E_a = 70$~MeV, the difference in $\sigma_{\rm el}$ between the choices of 
$r_A = 0.56$~\AA\,and $6.0$~\AA---see Eqs.~\eqref{eq:r2sc}
and \eqref{eq:r2r}---is within $20$\%, which is drastically small compared with the orders-of-magnitude difference seen in Fig.~\ref{fig:elcross}.

Next, we estimate the significance of the inelastic scattering processes for energetic axions. To that end, we first note that the incoherent scattering function $S(\bm{q}, Z)$ in Eq.~\eqref{eq:incscf} can be rewritten in the form 
\begin{align}
  S(\bm{q}, Z) &= \sum_{n} \biggl| \sum_i \langle A_n |  e^{-i \bm{q}\cdot \bm{x}_i } |A_0\rangle  \biggr|^2
  - \biggl| \sum_i \langle A_0 |  e^{-i \bm{q}\cdot \bm{x}_i } |A_0\rangle  \biggr|^2 \nonumber \\ 
  &= \langle A_0 | \sum_{i, j} e^{-i \bm{q}\cdot (\bm{x}_i-\bm{x}_j) } |A_0\rangle - |F(\bm{q})|^2 \nonumber \\ 
  &= Z - |F(\bm{q})|^2 +\langle A_0 | \sum_{i\neq j} e^{-i \bm{q}\cdot (\bm{x}_i-\bm{x}_j) } |A_0\rangle ~.
\end{align}
The last two terms are suppressed for $q r_A \gg 1$, and thus $S(\bm{q}, Z) \simeq Z$ in this limit. By comparing this with the factor $|Z -F(\bm{q})|^2 \simeq Z^2$, we find that the inelastic scattering cross section is smaller than the elastic scattering cross section by a factor of $Z^{-1}$, and thus is always subdominant for ${\cal O}(10)$~MeV axions. 

To summarize, we can estimate with good accuracy the total cross section for the inverse Primakoff process of 10--100~MeV axions by just using the simple expression in Eq.~\eqref{eq:sigelhe} with $r_A = {\cal O}(1)$~\AA.

It is of interest to estimate the number of the inverse Primakoff scattering events of axions from a nearby supernova explosion in the future neutrino experiments, such as Hyper-Kamiokande~\cite{Abe:2018uyc}, DUNE~\cite{Acciarri:2015uup, Abi:2018dnh, Abi:2020kei}, and JUNO~\cite{An:2015jdp}. There are quite a few supernova progenitor candidates that are located within a few hundred parsecs from the earth, with the nearest one (Spica; $\alpha$ Virginis) being at the distance of $d_{\mathrm{SN}} = 77(4)$~pc; for a list of supernova progenitor candidates, see, \textit{e.g.}, Table A1 in Ref.~\cite{Mukhopadhyay:2020ubs}. Following the prescription in Ref.~\cite{Ge:2020zww}, which is based on the result in Ref.~\cite{Carenza:2019pxu}, we estimate the axion production rate in a supernova explosion with the average energy of axions $\langle E_a \rangle = 70$~MeV as 
\begin{equation}
  \dot{N}_a \simeq 2.2 \times 10^{74} ~\mathrm{s}^{-1} \times \biggl(\frac{m_N}{f_a}\biggr)^2 C_{N, \mathrm{eff}}^2 ~,
\end{equation}
where $m_N$ is the nucleon mass and $C_{N, \mathrm{eff}}$ is given as a function of the axion-nucleon couplings, as shown in Ref.~\cite{Ge:2020zww}. For the KSVZ model, for example, $C_{N, \mathrm{eff}} = 0.37$, which we use as a benchmark. For the target atom, we consider oxygen ($Z=8$) for Hyper-Kamiokande~\cite{Abe:2018uyc}.\footnote{The inclusion of the scattering with hydrogen atoms does not change our estimation within the uncertainty of the calculation, because of the small atomic number of hydrogen ($Z = 1$) compared to oxygen.  } For a fiducial mass of 187~kt water~\cite{Abe:2018uyc}, there are $N_{\mathrm{O}}\simeq 6.26 \times 10^{33}$ oxygen atoms, for which we can estimate the expected number of events using Eq.~\eqref{eq:sigelhe} as 
\begin{align}
  N_{\mathrm{event}} &= 
  \dot{N}_a \Delta t \times \frac{\sigma_{\mathrm{el}}}{4\pi d^2} \times N_{\mathrm{O}} \nonumber \\ 
  &\simeq 1 \times \biggl(\frac{3 \times 10^8~\mathrm{GeV}}{f_a}\biggr)^4 \biggl(\frac{C_{N,\mathrm{eff}}}{0.37}\biggr)^2 \biggl(\frac{g_{a\gamma\gamma} f_a}{\alpha/\pi}\biggr)^2 \biggl(\frac{\Delta t}{10~\mathrm{s}}\biggr) 
  \biggl(\frac{d_{\mathrm{SN}}}{100~\mathrm{pc}}\biggr)^{-2} ~,
\end{align}
where we take $r_A = 2$~\AA\, and  $E_a = 70$~MeV in Eq.~\eqref{eq:sigelhe}, but the result has little dependence on this choice. As we see, we may have $\mathcal{O}(1)$ inverse Primakoff events for axions from a nearby supernova in Hyper-Kamiokande. Although it may be challenging to detect this signal in the presence of a huge amount of neutrinos from the same supernova, a more detailed study on the search strategy for this process would certainly be worthwhile.

%%%%%%%%%%%%%%%%%%%%%%
\section{Conclusion}
%%%%%%%%%%%%%%%%%%%%%%

We have discussed the calculation of the scattering cross section of the inverse Primakoff process for axions, with particular attention to its dependence on atomic form factors. This dependence is most significant for axions with energies $\lesssim \mathcal{O}(1)$~keV, and thus it is important to use a precise form factor in this case. We, in particular, have shown that the previous computations for xenon that use a screened Coulomb potential overestimate the cross section by more than an order of magnitude. We have also found that inelastic scattering cross sections may play an important role especially for low-energy scatterings. The scattering of more energetic axions, such as supernova axions, is also discussed, and the scattering cross section in this case is shown to be less dependent on the atomic structure. Finally, we have estimated the expected number of the inverse Primakoff process events for axions from a nearby supernova and found that $\mathcal{O}(1)$ events are expected in the future neutrino experiments. Our observation motivates a more dedicated study on the search strategy for this process, which we defer to future work.

%%%%%%%%%%%%%%%%%%%%%%%%%%%%%%%%%%%%
\subsection*{Note Added}
%%%%%%%%%%%%%%%%%%%%%%%%%%%%%%%%%%%%

After this article appeared on arXiv, Ref.~\cite{Gao:2020wer} is revised to a new version, Ref.~\cite{Gao:2020wer_v4}, where the authors changed the fitting formula for the form factor into a new one with an additional parameter ${\cal N}$; $F_{\mathrm{sc, mod}}(q)=Z [1+(1-{\cal N})q^2 r_0^2] / [1+q^2 r_0^2]$ in our notation, with ${\cal N}=0.54$ and $r_0=3.13$ \AA$^{-1}$. This fitting formula offers a good approximation to Eq.~\eqref{eq:apprx} for $|\bm{q}|\lesssim 5$~\AA$^{-1}$ and, as a consequence, the elastic scattering cross section based on this modified form factor is found to be in good agreement with our result for $E_a \lesssim 8$~keV. 

%%%%%%%%%%%%%%%%%%%%%%%%%%%%%%%%%%%%
\section*{Acknowledgments}
%%%%%%%%%%%%%%%%%%%%%%%%%%%%%%%%%%%%
We thank Yasuhiro Kishimoto for valuable discussions.
This work is supported in part by the Grant-in-Aid for Innovative Areas (No.19H05810 [TA and KH], No.19H05802 [KH], No.19H04615 [TA], No.18H05542 [NN]), Scientific Research B (No.20H01897 [KH and NN]), and Young Scientists B (No.17K14270 [NN]).

%%%%%%%%%%%%%%%%%%%%%%%%%%%%%%%%%%%%%%%%%%%%%%%%%%%%%%%%%%%%%
\section*{Appendix: Selected list of atomic form factors}
\appendix
%%%%%%%%%%%%%%%%%%%%%%%%%%%%%%%%%%%%%%%%%%%%%%%%%%%%%%%%%%%%

%%%%%%%%%%%%%%%%%%%%%%%%%%%%%%%%%%%%%%%%%%%%%%%%%%%%%%%%%%%%%%%%%%%%%%%%%%%%%%%5
\begin{table}
  \centering
  \caption{Parameters for the analytical approximation~\eqref{eq:apprx} to form factors of a selected set of target atoms and ions, taken from Ref.~\cite{formfactor}. }
  %\vspace{3mm}
  {\footnotesize 
  \begin{tabular}{ccccccccccc}
    \hline\hline
  & $Z$ & $a_1$ & $b_1$ & $a_2$  & $b_2$ & $a_3$  & $b_3$ & $a_4$  & $b_4$ &$c$ \\ 
  \hline 
  Xe & 54 & $20.2933$& $3.92820$& $19.0298$& $0.34400$& $8.97670$ & $26.4659$ & $1.99000$ & $64.2658$ & $3.71180$ \\ 
  Ge & 32 & $16.0816$& $2.85090$& $6.37470$& $0.25160$& $3.70680$ & $11.4468$ & $3.68300$ & $54.7625$ & $2.13130$ \\ 
  Na$^+$ & 11 & $3.25650$& $2.66710$& $3.93620$& $6.11530$& $1.39980$ & $0.20010$ & $1.00320$ & $14.0390$ & $0.40400$ \\ 
  I$^-$ & 53 & $20.2332$& $4.35790$& $18.9970$& $0.38150$& $7.80690$ & $29.5259$ & $2.88680$ & $84.9304$ & $4.07140$ \\ 
  Pb & 82 & $31.0617$& $0.69020$& $13.0637$& $2.35760$& $18.4420$ & $8.61800$ & $5.96960$ & $47.2579$ & $13.4118$ \\ 
    \hline\hline
  \end{tabular}
  }
  \label{tab:formfactors}
\end{table}
%%%%%%%%%%%%%%%%%%%%%%%%%%%%%%%%%%%%%%%%%%%%%%%%%%%%%%%%%%%%%%%%%%%%%%%%%%%%%%%%%%%%%%

In Table~\ref{tab:formfactors}, we summarize the parameters of the approximated analytical function~\eqref{eq:apprx} for a selected choice of target atoms and ions used in experiments. These parameters are again taken from Ref.~\cite{formfactor}. For the choice of target atoms and ions (besides xenon, whose parameters are also presented in Sec.~\ref{sec:formfactor}), we follow Table~A.1 in Ref.~\cite{Cebrian:1998mu}.

%%%%%%%%%%%%%%%%%%%%%%%%%%%%%%%%%%%%%%%%%%%%%%%%%%%%%%%%%%%%%%%%
\begin{figure}
  \centering
  \subcaptionbox{\label{fig:formfactor_ge} Ge}{
  \includegraphics[width=0.45\columnwidth]{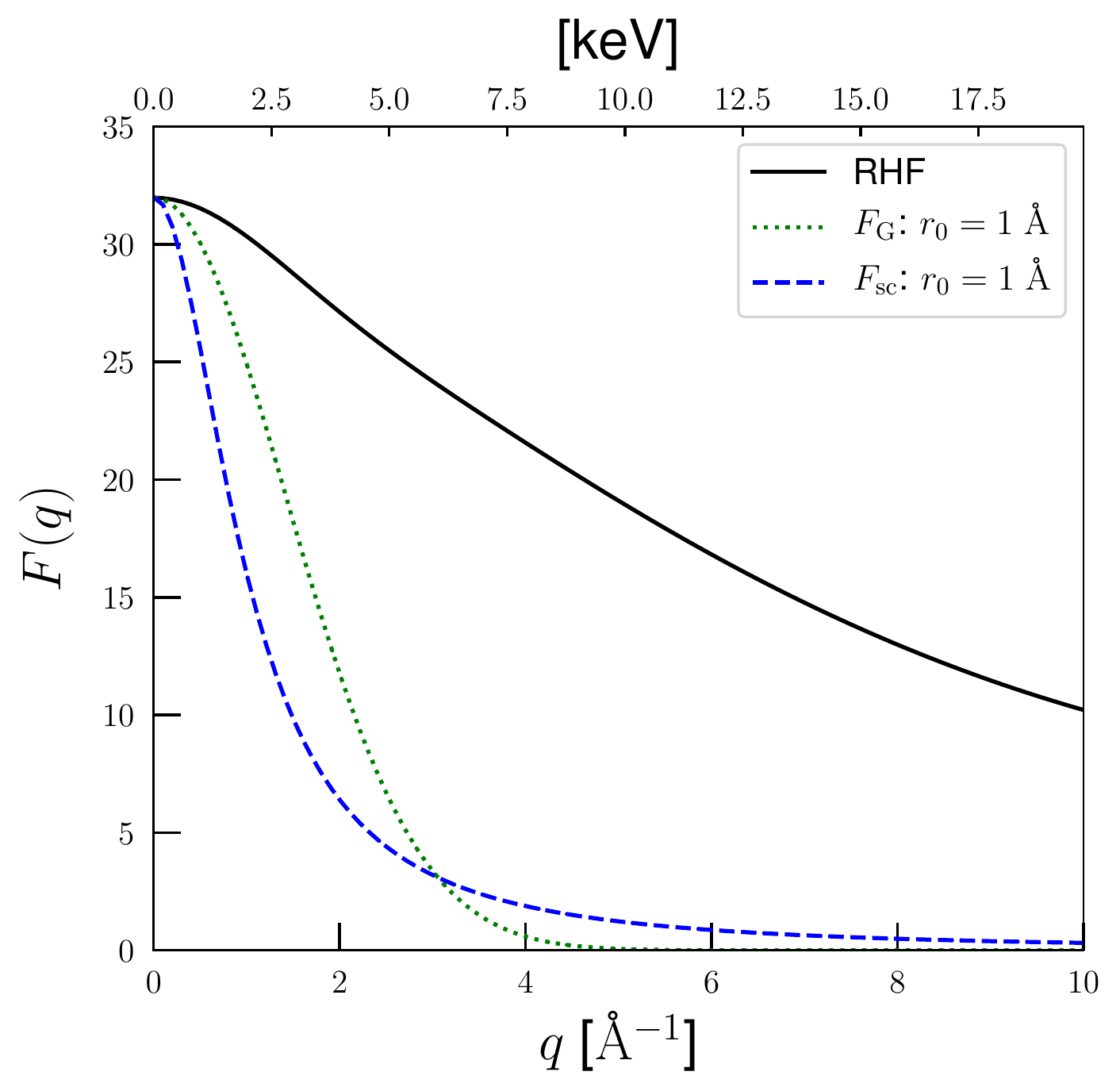}}
 \hspace{5mm}
  \subcaptionbox{\label{fig:formfactor_pb} Pb}{
  \includegraphics[width=0.45\columnwidth]{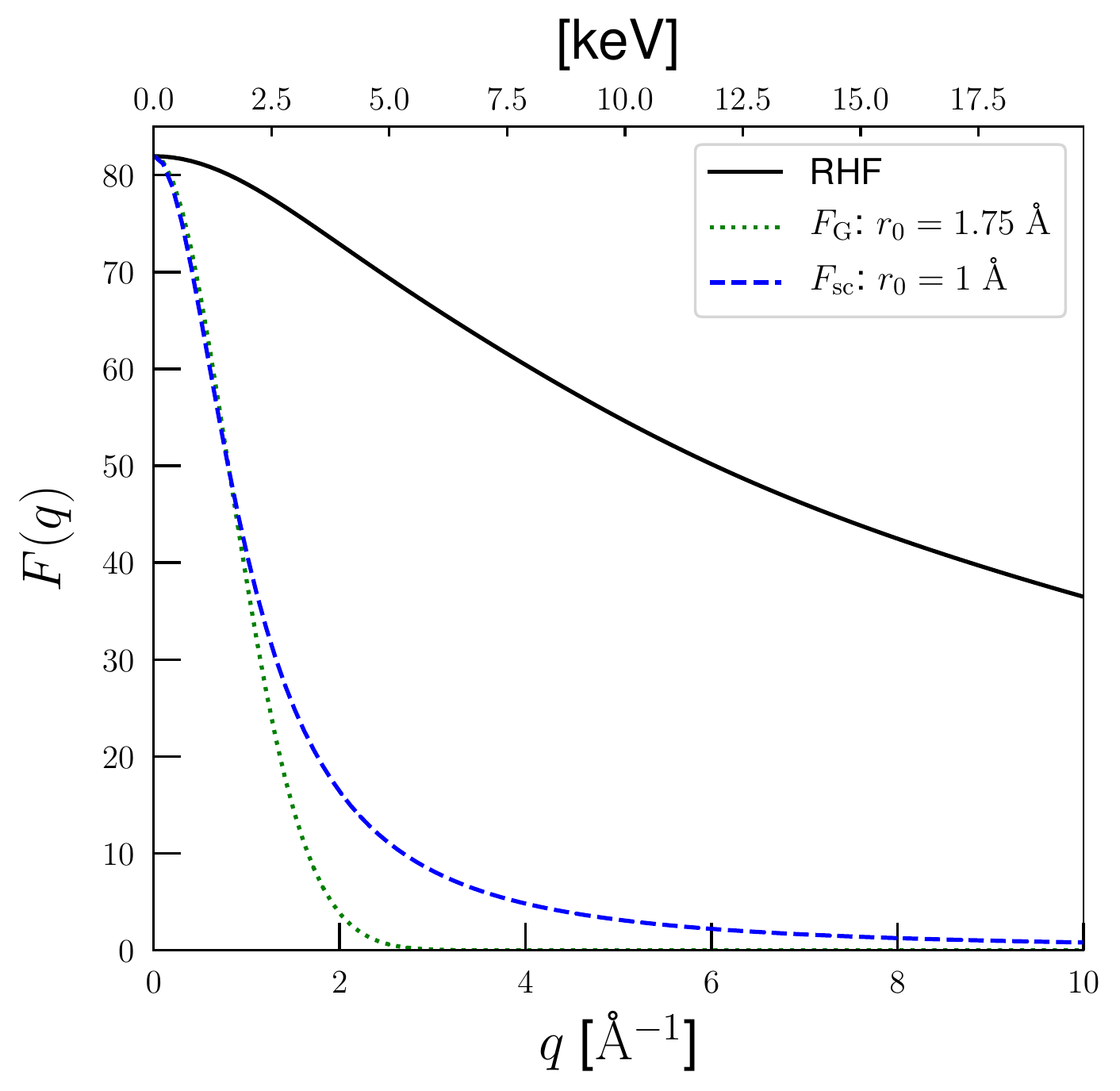}}
  \caption{Atomic form factors of (a) Ge and (b) Pb as functions of momentum transfer $q = |\bm{q}|$. The black solid, blue dashed, and green dotted lines correspond to the form factor obtained with a relativistic Hartree-Fock wave function (we use the analytical approximation~\eqref{eq:apprx} with the parameters in Table~\ref{tab:formfactors}), $F_{\rm sc}(q; r_0)$, and $F_G (q; r_0)$, respectively. For $F_{\rm sc}(q; r_0)$, the screening length is set to be $r_0 = 1$~{\AA} as in Ref.~\cite{Cebrian:1998mu}, while for $F_G (q; r_0)$, we set $r_0 = 1$~{\AA} and 1.75~{\AA} for Ge and Pb, respectively~\cite{Avignone:1988bv}.}
  \label{fig:formfactors2}
  \end{figure}
  %%%%%%%%%%%%%%%%%%%%%%%%%%%%%%%%%%%%%%%%%%%%%%%%%%%%%%%%%%%%%%

In addition to $F_{\rm sc}(q; r_0)$ in Eq.~\eqref{eq:fsc}, a Gaussian form factor,  
\begin{equation}
  F_G (q; r_0) \equiv Z e^{-q^2 r_0^2/4} ~,
\end{equation}
was used in the literature~\cite{Avignone:1988bv}. To compare this form factor with those discussed in the main text, in Fig.~\ref{fig:formfactors2}, we show the atomic form factors of (a) Ge and (b) Pb as functions of momentum transfer $q = |\bm{q}|$. The black solid, blue dashed, and green dotted lines correspond to the form factor obtained with a relativistic Hartree-Fock wave function (we use the analytical approximation~\eqref{eq:apprx} with the parameters in Table~\ref{tab:formfactors}), $F_{\rm sc}(q; r_0)$, and $F_G (q; r_0)$, respectively. For $F_{\rm sc}(q; r_0)$, the screening length is set to be $r_0 = 1$~{\AA} as in Ref.~\cite{Cebrian:1998mu}, while for $F_G (q; r_0)$, we set $r_0 = 1$~{\AA} and 1.75~{\AA} for Ge and Pb, respectively~\cite{Avignone:1988bv}. These plots show that both of the form factors are smaller than that computed with a relativistic Hartree-Fock wave function for ${\cal O}(1)$~keV momentum transfer. As a result, the cross sections of the inverse Primakoff scattering for these atoms were again overestimated in the previous studies.

%%%%%%%%%%%%%%%%%%%%%%%%%%%%%%%%%%%%%%%

\bibliographystyle{utphysmod}
\bibliography{ref}

%%%%%%%%%%%%%%%%%%%%%%%%%%%%%%%%%%%%%%%

\end{document}